\documentclass[12pt]{article}

\usepackage{amssymb}
\usepackage{amscd}

\newcommand{\dd}{\mathrm{d}}

\newcommand{\vect}{\mathbf}
\newcommand{\Lied}[1]{\mathcal{L}_{\vect{#1}}}

\newcommand{\XS}{\mathcal{X}_{\mathrm{S}}}
\newcommand{\XH}{\mathcal{X}_{\mathrm{H}}}
\newcommand{\tr}{\mathrm{tr\,}}

\newtheorem{thm}{Theorem}
\newtheorem{corol}{Corollary}
\newtheorem{lem}{Lemma}

\newcommand{\comments}[1]{{}}
\newcommand{\tedious}[1]{}
\newcommand{\definition}{\emph}

\title{General Volume-Preserving Mechanical Systems}
\author{Bin Zhou\thanks{Email: zhoub@ihep.ac.cn}\\
  Institute of High Energy Physics, Chinese Academy of Sciences, \\
  P. O. Box 918-4, Beijing 100039, P. R. China, \\
       Han-Ying Guo\thanks{hyguo@itp.ac.cn}\\
  Institute of Theoretical Physics, Chinese Academy of Sciences, \\
  P. O. Box 2735, Beijing 100080, P. R. China, \\
       and Ke Wu\thanks{wuke@itp.ac.cn}\\
  Department of Mathematics, Capital Normal University, \\
  Beijing 100037, P. R. China}

\date{July 18, 2002}

\begin{document}

\maketitle

\begin{abstract}
In this letter, we present the general form of equations that generate a
volume-preserving flow on a symplectic manifold $(\mathcal{M, \omega})$.
It is shown that every volume-preserving flow has some 2-forms acting the
r\^ole of the Hamiltonian functions in the Hamiltonian mechanics and the
ordinary Hamilton equations are included as a special case with a 2-form
$\frac{1}{n-1}\,H\,\omega$ where $H$ is the corresponding Hamiltonian.
\end{abstract}

\comments{
}

In classical mechanics, most of the systems can be described by the ordinary
Hamilton equations
\begin{equation}
  \dot{q}^{\,i} = \frac{\partial H}{\partial p_i}, \qquad
  \dot{p}_i = - \frac{\partial H}{\partial q^i}.
\label{Heqs}
\end{equation}
For an autonomous system, namely, a system with the Hamiltonian manifestly
independent of the time $t$, the Hamiltonian $H$ is a function on the phase
space $\mathcal{M}$ which is precisely the cotangent bundle of the
configuration space.  Abstractly, a Hamiltonian system can be defined on a
symplectic manifold $(\mathcal{M},\omega)$ where $\mathcal{M}$ is a
$2n$-dimensional orientable differentiable manifold endowed with a symplectic
2-form $\omega$  at every point of $\mathcal{M}$. For the details of
Hamiltonian mechanics on a symplectic manifold, see, for example,
\cite{Arnold}. Then, according to Darboux's theorem, for each point $x$ of
$\mathcal{M}$, there is a coordinate neighborhood
  $(U; q^1, \dots,q^n,p_1,\dots,p_n)$ of $x$ such
that\footnote{
Repeated indices are always assumed to be summed over 1 to $n$ unless otherwise
stated. For convenience, symbols such like $\omega|_{U}$ are always abbreviated
as $\omega$.
}
\begin{equation}
  \omega|_{U} = \dd p_i\wedge\dd q^i.
\label{omegadef}
\end{equation}
Thus a Hamiltonian system with the Hamiltonian $H$ can still be described
locally by eqs.~(\ref{Heqs}). The solutions of eqs.~(\ref{Heqs}) are the
integral curves of the Hamiltonian vector field
\begin{equation}
  \vect{X}_H := \frac{\partial H}{\partial p_i}\frac{\partial}{\partial q^i}
  - \frac{\partial H}{\partial q^i}\frac{\partial}{\partial p_i}
\end{equation}
on $\mathcal{M}$. It is well known that the above ordinary Hamiltonian system
preserves both the symplectic 2-form $\omega$ and the volume form
$\frac{1}{n!}\, \omega^n$,
where $n = \frac{1}{2}\, \dim\mathcal{M}$, and
$\omega^k$ is the $k$-fold wedge product of $\omega$.

It is well known that the ordinary Hamilton equations can be expressed as
\begin{equation}
  - i_{\vect{X}_H}\omega = \dd H,
\end{equation}
where $i$ denotes the contraction. It is easy to show the symplectic
structure-preserving and volume-preserving properties for these ordinary
Hamiltonian systems. Namely,
\begin{equation}
  \Lied{X_{\mathnormal{H}}}\omega = 0
\end{equation}
and
\begin{equation}
  \Lied{X_{\mathnormal{H}}}(\omega^n)
  = n\,(\Lied{X_{\mathnormal{H}}}\omega)\wedge\omega^{n-1}=0,
\end{equation}
respectively.

However, there exist certain important mechanical systems that preserve the
volume of the phase space only and cannot be transformed into the ordinary
Hamiltonian systems. For example, the linear systems on $\mathbb{R}^{n}$
\begin{equation}
  \ddot{q}^{\ i} = - k_{ij} \,q^j
\label{ls}
\end{equation}
with constant coefficients $k_{ij}$ are such kind of mechanical systems.
Obviously, eqs. (\ref{ls}) can be turned into the form:
\begin{equation}
  \dot{q}^{\ i} = \frac{\partial H}{\partial p_i}, \qquad
  \dot{p}_i = - \frac{\partial H}{\partial q^i} - a_{ij}\, q^j
\label{lsys}
\end{equation}
with
\begin{eqnarray}
  & & H = \frac{1}{2}\,\delta^{ij}\,p_i p_j
  + \frac{1}{2}\, s_{ij}\,q^i q^j,
\label{lsysH} \\
  & & s_{ij} = \frac{1}{2}\,( k_{ij} + k_{ji}), \qquad
  a_{ij} = \frac{1}{2}\,(k_{ij} - k_{ji}).
\label{lsyscoef}
\end{eqnarray}
When one of $a_{ij}$ is nonzero\footnote{
Here is an example showing that even a conservative system can be transformed to
be a non-Hamiltonian system. Consider a system consisting of two 1-dimensional
linearly coupled oscillators:
\begin{displaymath}
  m_1\,\ddot{q}^{\,1} = -k\,(q^2 - q^1), \qquad
  m_2\,\ddot{q}^{\,2} = -k\,(q^1 - q^2).
\end{displaymath}
Obviously, such a system satisfies Newton's laws. Let
$k_{11} = -k/m_1$, $k_{12} = k/m_1$, $k_{21} = k/m_2$ and $k_{22} = -k/m_2$.
Then it is a system as described by eqs.~(\ref{ls}), with the indices running
from 1 to 2. When $m_1 \neq m_2$, we have a system satisfying
$k_{12} \neq k_{21}$.
},
the system (\ref{lsys}) is not an ordinary Hamiltonian system on
$\mathbb{R}^{2n}$.  In fact, the corresponding vector field
\begin{displaymath}
  \vect{X} = \vect{X}_H - a_{ij}\,q^j\,\frac{\partial}{\partial p_i}
\end{displaymath}
of (\ref{lsys}) is not even a symplectic vector field since the Lie derivative
\begin{displaymath}
  \Lied{X}\omega = a_{ij}\,\dd q^i\wedge\dd q^j
\end{displaymath}
does not vanish in this case. On the other hand, the linear system (\ref{lsys})
always preserves the volume form of the phase space because
$\Lied{X}(\omega^n) = 0$, as what can be easily verified.

In this letter, we present a general form of
equations that generate a volume-preserving flow
on a symplectic manifold $(\mathcal{M, \omega})$.
It is shown that every volume-preserving flow has
some 2-forms acting the r\^ole of the Hamiltonian
functions in the Hamiltonian mechanics and the
ordinary Hamiltonian equations are included as a
special case with a 2-form
$\frac{1}{n-1}\,H\,\omega$ where $H$ is the
corresponding Hamiltonian.

The main idea for presenting such general
equations for volume-preserving systems is based
on the linear isomorphism between the space
$\XH^n(\mathcal{M},\omega)$ of certain
volume-preserving vector fields and the space
$B^{2n-1}(\mathcal{M})$ of exact $(2n-1)$-forms
on $\mathcal{M}$. This is, in fact, a direct
generalization for the ordinary Hamiltonian
system ($n = 1$). Namely, the linear isomorphism
between the space $\XH^{n=1}(\mathcal{M},\omega)$
of certain symplectic structure-preserving vector
fields and the space $B^{1}(\mathcal{M})$ of
exact $1$-forms on $\mathcal{M}$.

Briefly speaking, a volume-preserving system,
described by a volume-preserving vector field
with respect to which the Lie derivative of
$\omega^n$ being zero, corresponds to a unique
closed $(2n-1)$-form on $\mathcal{M}$ and
\textit{vise versa}. Such a system that
corresponds to an exact $(2n-1)$-form can be
determined then by a 2-form (may up to a closed
2-form), as if a Hamiltonian system can be
determined by a function (the Hamiltonian) up to
a constant function.

In this letter, the main results will be quoted and outlined without detailed
proof. For the details and extended topics, we refer to our forthcoming papers.

\vskip 10mm

For each integer $k = 1,\ldots,n$ where $n =
\frac{1}{2}\,\dim\mathcal{M} \geqslant 1$, we may define the
set of the symplectic vector fields and that of
the Hamiltonian vector fields as
\begin{eqnarray*}
  \XS^k(\mathcal{M},\omega) & = & \{\,\vect{X}\in\mathcal{X(M)}\,|\,
  \Lied{X}(\omega^k) = 0 \,\}, \\
  \XH^k(\mathcal{M},\omega) & = & \{\,\vect{X}\in\mathcal{X(M)}\,|\,
  - i_{\vect{X}}(\omega^k) \textrm{ is exact}\},
\end{eqnarray*}
respectively. Here and hereafter $\mathcal{X(M)}$ denotes the space of smooth
vector fields on $\mathcal{M}$. Then it is easy to prove that
\begin{eqnarray*}
  & & \XS^1(\mathcal{M},\omega) \subseteq \ldots \subseteq
  \XS^k(\mathcal{M},\omega) \subseteq \XS^{k+1}(\mathcal{M},\omega)
  \subseteq \ldots \subseteq\XS^n(\mathcal{M},\omega), \\
  & & \XH^1(\mathcal{M},\omega) \subseteq \ldots \subseteq
  \XH^k(\mathcal{M},\omega) \subseteq \XH^{k+1}(\mathcal{M},\omega)
  \subseteq \ldots \subseteq\XH^n(\mathcal{M},\omega)
\end{eqnarray*}
and
\begin{displaymath}
  \XH^k(\mathcal{M},\omega) \subseteq \XS^k(\mathcal{M},\omega).
\end{displaymath}
In fact, $\XS^k(\mathcal{M},\omega)$ is a Lie algebra under the commutation
bracket of vector fields, and $\XH^k(\mathcal{M},\omega)$ is an ideal of
$\XS^k(\mathcal{M},\omega)$ because
\begin{displaymath}
  [\XS^k(\mathcal{M},\omega),\XS^k(\mathcal{M},\omega)]
  \subseteq \XH^k(\mathcal{M},\omega).
\end{displaymath}
Let us consider the quotient Lie algebra
\begin{equation}
  H^k_{\mathrm{EL}}(\mathcal{M},\omega)
  :=
  \XS^k(\mathcal{M},\omega)/\XH^k(\mathcal{M},\omega).
\end{equation}
It can be  called the $k$-th
\definition{Euler-Lagrange cohomology group}. It
is called a group because its Lie algebra
structure is trivial: For each $k = 1$, $\ldots$,
$n$, $H^k_{\mathrm{EL}}(\mathcal{M},\omega)$ is
an abelien Lie algebra. For $n=1$,
$H^1_{\mathrm{EL}}(\mathcal{M},\omega)$ is the
Euler-Lagrange cohomology first introduced in
\cite{ELcoh2,ELcoh4}. And it can be proved that
$H^1_{\mathrm{EL}}(\mathcal{M},\omega)$ is
linearly isomorphic to the first de~Rham
cohomology group
$H^1_{\mathrm{dR}}(\mathcal{M})$.

For $k = n$, $\XS^n(\mathcal{M},\omega)$ is the
Lie algebra of volume-preserving vector fields,
including all the other Lie algebras mentioned in
the above as its Lie subalgebras. Similarly to
the case of $k = 1$, the $n$-th Euler-Lagrange
cohomology group
$H^n_{\mathrm{EL}}(\mathcal{M},\omega)$ is
isomorphic to the $(2n - 1)$-th de~Rham
cohomology group
$H^{2n-1}_{\mathrm{dR}}(\mathcal{M})$. To prove
this, let us first introduce a lemma:

\begin{lem}
  For each integer $k=1,\ldots, n$ $(n \geqslant 1)$ and $x\in\mathcal{M}$, the
linear map $\nu_k: T_x \mathcal{M} \longrightarrow \Lambda_{2k-1}(T^*_x
\mathcal{M})$ defined by $\nu_k(\vect{X}) = - i_{\vect{X}}(\omega^k)$ is
injective, namely, $i_{\vect{X}}(\omega^k) = 0$ iff $\vect{X} = 0$.
\label{lem:lemma1}
\end{lem}

Since $\dim T_x\mathcal{M} = \dim \Lambda_{2n-1}(T^*_x\mathcal{M})$, lemma
\ref{lem:lemma1} implies that $\nu_n: T_x\mathcal{M}\longrightarrow
\Lambda_{2n-1}(T^*_x\mathcal{M})$ is a linear isomorphism, which induces a
linear isomorphism
$\nu_n:\mathcal{X(M)}\longrightarrow\Omega^{2n-1}(\mathcal{M})$
defined by
\begin{equation}
  \nu_n(\vect{X}) = - i_{\vect{X}}(\omega^n)
  = -n\,(i_{\vect{X}}\omega)\wedge\omega^{n-1}.
\end{equation}
Hence we obtain
\begin{thm}
The linear map $\nu_n: \mathcal{X(M)}\longrightarrow\Omega^{2n-1}(\mathcal{M})$
is an isomorphism. Under this isomorphism,
$\XS^n(\mathcal{M},\omega)$ and $\XH^n(\mathcal{M},\omega)$ are isomorphic to
$Z^{2n-1}(\mathcal{M})$ and $B^{2n-1}(\mathcal{M})$, respectively.
\label{thm:thm1}
\end{thm}
In the above theorem, $Z^{2n-1}(\mathcal{M})$ is the space of closed
$(2n-1)$-forms on $\mathcal{M}$ and $B^{2n-1}(\mathcal{M})$ is the space of
exact $(2n-1)$-forms on $\mathcal{M}$. As a consequence, we obtain the
following corollary:
\begin{corol}
  The $n$-th Euler-Lagrange cohomology group
$H^n_{\mathrm{EL}}(\mathcal{M},\omega)$ is linearly isomorphic to
$H^{2n-1}_{\mathrm{dR}}(\mathcal{M})$, the $(2n-1)$-th de~Rham cohomology group.
\end{corol}

When $\mathcal{M}$ is closed (i.e., compact and
without boundary), such as a closed Riemann surface,
$H^n_{\mathrm{EL}}(\mathcal{M},\omega)$ is
linearly isomorphic to the dual space of
$H^1_{\mathrm{EL}}(\mathcal{M})$, because
  $H^k_{\mathrm{dR}}(\mathcal{M}) \cong (H^{2n-k}_{\mathrm{dR}}(\mathcal{M}))^*$
for such a manifold (see, for example, \cite{Warner}).
If $\mathcal{M}$ is not compact, this relation cannot be assured.

When $n > 2$, there is another lemma:
\begin{lem}
If $n>2$, then, for an arbitrary integer $k = 1, \ldots, n-2$ and a
point $x\in\mathcal{M}$, $\alpha\in\Lambda_2(T^*_x\mathcal{M})$ satisfies
$\alpha\wedge\omega^k = 0$ iff $\alpha = 0$.
\label{lem:lemma2}
\end{lem}
Again this induces injective linear maps
\begin{eqnarray*}
  \iota_k: \Omega^2(\mathcal{M}) & \longrightarrow & \Omega^{2k+2}(\mathcal{M})
\\
  \alpha & \longmapsto & \alpha\wedge\omega^k
\end{eqnarray*}
for $k = 1,\ldots, n-2$. Thus we can obtain that
\begin{displaymath}
  \XS^1(\mathcal{M},\omega) = \XS^2(\mathcal{M},\omega) = \ldots
  = \XS^{n-1}(\mathcal{M},\omega) \subseteq \XS^n(\mathcal{M},\omega).
\end{displaymath}
When $k = n-2$, it can be easily verified from lemma \ref{lem:lemma2} that
\begin{eqnarray}
  \iota = \iota_{n-2}: \Omega^2(\mathcal{M}) & \longrightarrow &
  \Omega^{2n-2}(\mathcal{M})
\nonumber \\
  \alpha & \longmapsto & \alpha\wedge\omega^{n-2}
\end{eqnarray}
is a linear isomorphism. If $n = 2$, we use the convention that $\omega^0 = 1$,
namely, $\iota = \mathrm{id}: \alpha \longmapsto \alpha$.
Then we can define a linear map $\phi$ making the following diagram commutative:
\begin{equation}
  \begin{CD}
    \Omega^2(\mathcal{M}) @>\iota>> \Omega^{2n-2}(\mathcal{M}) \\
    @V\phi VV                        @V\dd VV\\
    \XH^n(\mathcal{M},\omega) @>\nu_n>> B^{2n-1}(\mathcal{M})
  \end{CD}
\label{phicg}
\end{equation}
Equivalently, given a 2-form
\begin{equation}
  \alpha = \frac{1}{2}\,Q_{ij}\,\dd q^i\wedge\dd q^j
  + A^i_j\,\dd p_i\wedge\dd q^j + \frac{1}{2}\,P^{ij}\,\dd p_i\wedge\dd p_j
\label{alpha}
\end{equation}
where $Q_{ij}$ and $P^{ij}$ satisfy
\begin{equation}
  Q_{ji} = - Q_{ij}, \qquad P^{ji} = - P^{ij},
\end{equation}
the vector field
\begin{displaymath}
  \phi(\alpha) = (\nu_n^{-1}\circ\dd\circ\iota)(\alpha)
  = \nu^{-1}_n(\dd\alpha\wedge\omega^{n-2})
\end{displaymath}
is in $\XH^n(\mathcal{M},\omega)$.

For convenience, we set
\begin{displaymath}
  \vect{X} = n(n-1)\,\phi(\alpha)
  = n(n-1)\,\nu_n^{-1}(\dd\alpha\wedge\omega^{n-2}),
\end{displaymath}
namely, $i_{\vect{X}}(\omega^n) = -
n(n-1)\,(\dd\alpha)\wedge\omega^{n-2}$. It is
easy to obtain that
\begin{equation}
  \vect{X} = \bigg(\frac{\partial P^{ij}}{\partial q^j}
  + \frac{\partial A^j_j}{\partial p_i} - \frac{\partial A^i_j}{\partial p_j}
  \bigg)\,\frac{\partial}{\partial q^i}
  + \bigg(\frac{\partial Q_{ij}}{\partial p_j}
  - \frac{\partial A^j_j}{\partial q^i}
  + \frac{\partial A^j_i}{\partial q^j}\bigg)\,
  \frac{\partial}{\partial p_i}.
\label{XH}
\end{equation}
 In fact,
\begin{eqnarray*}
  \dd\iota(\alpha)&=&\dd\alpha\wedge\omega^{n-2} \\
  &=& \frac{1}{2}\,\frac{\partial Q_{ij}}{\partial q^k}\,
  \dd q^i\wedge\dd q^j\wedge\dd q^k\wedge\omega^{n-2}
  +\frac{1}{2}\,\frac{P^{ij}}{\partial p_k}\,
    \dd p_i\wedge\dd p_j\wedge\dd p_k\wedge\omega^{n-2}
\\
  & & +\bigg(\frac{1}{2}\,\frac{\partial Q_{jk}}{\partial p_i}
    + \frac{\partial A^i_j}{\partial q^k}\bigg)\,
    \dd p_i\wedge\dd q^j\wedge\dd q^k\wedge\omega^{n-2}
\\
  & & + \bigg( \frac{1}{2}\,\frac{\partial P^{ij}}{\partial q^k}
    -\frac{\partial A^i_k}{\partial p_j} \bigg)\,
    \dd p_i\wedge\dd p_j\wedge\dd q^k\wedge\omega^{n-2}
\\
  &=&  \bigg(\frac{1}{2}\,\frac{\partial Q_{jk}}{\partial p_i}
    + \frac{\partial A^i_j}{\partial q^k}\bigg)\,
    \dd p_i\wedge\dd q^j\wedge\dd q^k\wedge\omega^{n-2}
\\
  & & + \bigg( \frac{1}{2}\,\frac{\partial P^{ij}}{\partial q^k}
    -\frac{\partial A^i_k}{\partial p_j} \bigg)\,
    \dd p_i\wedge\dd p_j\wedge\dd q^k\wedge\omega^{n-2}.
\end{eqnarray*}
By virtue of the following two equations
\begin{eqnarray}
  \dd p_i\wedge\dd p_j\wedge\dd q^k \wedge\omega^{n-2}
  & = & \frac{\delta^k_j}{n-1}\,\dd p_i \wedge\omega^{n-1}
  - \frac{\delta^k_i}{n-1}\,\dd p_j \wedge\omega^{n-1},
\\
  \dd p_i\wedge\dd q^j\wedge\dd q^k\wedge\omega^{n-2}
  & = & \frac{\delta^j_i}{n-1}\,\dd q^k\wedge\omega^{n-1}
  - \frac{\delta^k_i}{n-1}\,\dd q^j\wedge\omega^{n-1},
\end{eqnarray}
we can write $\dd\iota(\alpha)$ as
\begin{eqnarray*}
  \dd\iota(\alpha)
  & = & \frac{1}{n-1}\bigg(\frac{\partial A^j_j}{\partial q^i}
  - \frac{\partial A^j_i}{\partial q^j} - \frac{\partial Q_{ij}}{\partial p_j}
  \bigg)\,\dd q^i\wedge\omega^{n-1}
\nonumber \\ & &
  + \frac{1}{n-1}\bigg(\frac{\partial P^{ij}}{\partial q^j}
  + \frac{\partial A^j_j}{\partial p_i} - \frac{\partial A^i_j}{\partial p_j}
  \bigg)\,\dd p_i\wedge\omega^{n-1}
\\
  & = & \frac{1}{n(n-1)}\bigg(\frac{\partial A^j_j}{\partial q^i}
  - \frac{\partial A^j_i}{\partial q^j} - \frac{\partial Q_{ij}}{\partial p_j}
  \bigg)\,i_{\frac{\partial}{\partial p_i}}\omega^n
\nonumber \\ & &
  - \frac{1}{n(n-1)}\bigg(\frac{\partial P^{ij}}{\partial q^j}
  + \frac{\partial A^j_j}{\partial p_i} - \frac{\partial A^i_j}{\partial p_j}
  \bigg)\,i_{\frac{\partial}{\partial q^i}}\omega^n.
\end{eqnarray*}
Comparing it with
\begin{equation}
  \dd\iota(\alpha) = \frac{1}{n(n-1)}\,\nu_n(\vect{X})
  = -\frac{1}{n(n-1)}\,i_{\vect{X}}(\omega^n),
\label{nunX}
\end{equation}
we can obtain the expression of $\vect{X}$, as shown in eq.~(\ref{XH}).

Since both $\iota$ and $\nu_n$ are linear
isomorphisms, we can see from the commutative
diagram (\ref{phicg}) that for each 2-form
$\alpha$ on $\mathcal{M}$ as in eq.~(\ref{alpha})
the vector field $\vect{X}$ in eq.~(\ref{XH})
belongs to $\XH^n(\mathcal{M},\omega)$ and for
each
  $\vect{X}\in\XH^n(\mathcal{M},\omega)$
there exists the 2-form $\alpha$ on $\mathcal{M}$ satisfying eqs.~(\ref{XH}).
But there may be several 2-forms that are mapped to the same vector field
$\vect{X}$. For example, the vector field $\vect{X}$ in (\ref{XH}) is invariant
under the transformation
\begin{equation}
  \alpha \longmapsto \alpha + \theta
\label{transf}
\end{equation}
where $\theta$ is a closed 2-form.

Note that for the vector field
$\vect{X}\in\XH^n(\mathcal{M},\omega)$, the
2-form $\alpha$ is a globally defined
differential form on $\mathcal{M}$. If
$\vect{X}\in\XS^n(\mathcal{M},\omega)$, such a
2-form cannot be found if
$H^{2n-1}_{\mathrm{dR}}(\mathcal{M})$ is
nontrivial. In this case, $\alpha$ can be still
found as a locally defined 2-form, according to
Poincar\'e's lemma. Then the relation between
$\vect{X}$ and the locally defined 2-form
$\alpha$, eq.~(\ref{XH}), is valid only on a
certain  open subset of $\mathcal{M}$, and, on
the intersection of two such open subsets, the
corresponding 2-forms are not identical. That is,
the transformation relation of $Q_{ij}$, $A^i_j$
as well as $P^{ij}$ is not that of tensors on
$\mathcal{M}$. Instead, a transformation such
like (\ref{transf}) or more complicated should be
applied.

No matter whether $\vect{X}$ belongs to $\XH^n(\mathcal{M},\omega)$ or
$\XS^n(\mathcal{M},\omega)$, namely, whether the 2-form $\alpha$ is globally or
locally defined, the flow of $\vect{X}$ can be always obtained provided that
the general solution of the following equations can be solved:
\begin{eqnarray}
  \dot{q}^{\,i} & = & \frac{\partial P^{ij}}{\partial q^j}
  + \frac{\partial A^j_j}{\partial p_i}
  - \frac{\partial A^i_j}{\partial p_j},
\nonumber \\
  \dot{p}_i & = & \frac{\partial Q_{ij}}{\partial p_j}
  - \frac{\partial A^j_j}{\partial q^i}
  + \frac{\partial A^j_i}{\partial q^j}.
\label{evps}
\end{eqnarray}
This is just the general form of the equations of
a volume-preserving mechanical system on a
symplectic manifold $(\mathcal{M},\omega)$.

For a function $H$ on $\mathcal{M}$, let the 2-form be
\begin{equation}
  \alpha = \frac{1}{n-1}\,H\,\omega,
\end{equation}
then eqs.~(\ref{evps}) turn to be the ordinary Hamilton equations
(\ref{Heqs}). Thus the ordinary Hamilton equations have been included
as a special case, as what is expected.

Now suppose the symplectic manifold is $\mathbb{R}^{2n}$ with the standard
symplectic form $\omega=\dd p_i \wedge\dd q^i$. When
  $Q_{ij} = - a_{ij}\,p_k q^k$, $A^i_j = \frac{1}{n-1}\,H\,\delta^i_j$
and $P^{ij} = 0$
with the constants $a_{ij}$ and the function $H$ as shown in eqs.~(\ref{lsysH})
and (\ref{lsyscoef}), namely,
\begin{equation}
  \alpha = \frac{1}{n-1}\,H\,\omega
  - \frac{1}{2}\,p_k q^k\,a_{ij}\,
  \dd q^i\wedge\dd q^j,
\label{lsysalpha}
\end{equation}
then the equations (\ref{evps}) will turn into
eqs.~(\ref{lsys}). Therefore the 2-form $\alpha$
in eq.~(\ref{lsysalpha}) is one of general forms
corresponding to a linear system, but it is not
the unique.

It is interesting that when $H = 0$ in the above equation, all the coordinates
$q^i$ are the first integrals of the linear system. Then all the canonical
momenta
  $p_i = a_{ij}\,q^j t + p_{i0}$ where $p_{i0}$
are constants. This can generalize to an arbitrary symplectic manifold for the
equations (\ref{evps}), even though the system is no longer a linear system:
If, on a Darboux coordinate neighborhood $(U; q,p)$ in a symplectic manifold
$(\mathcal{M},\omega)$, the 2-form $\alpha$ satisfies
\begin{equation}
  i_{\frac{\partial}{\partial p_i}}\alpha
  = A^i_j\,\dd q^j + P^{ij}\,\dd p_j
  = 0
\end{equation}
for each $i=1,\ldots,n$, then $A^i_j = 0$ and $P^{ij} = 0$. Hence, on that
coordinate neighborhood $U$, all the $q^i$ are constant.

If we define a function $\tr\alpha$ as
\begin{equation}
  \alpha\wedge\omega^{n-1} = \frac{\tr\alpha}{n} \,\omega^n
\end{equation}
for each 2-form $\alpha$, then using the formula
\begin{displaymath}
  \dd p_i\wedge\dd q^j\wedge\omega^{n-1} = \frac{\delta^j_i}{n}\,\omega^n,
\end{displaymath}
we obtain that
\begin{equation}
  \tr\alpha = A^i_i.
\end{equation}
The above expression is obviously independent of the choice of the Darboux
coordinates. Let $\vect{X}_{\tr\alpha}$ be the Hamiltonian vector field
corresponding to the function $\tr\alpha$. Eq.~(\ref{XH}) indicates that
\begin{equation}
  \vect{X} = \vect{X}_{\tr\alpha} + \vect{X}'
\end{equation}
where $\vect{X}'$ is the extra part on the right hand side of eq.~(\ref{XH}),
corresponding to the 2-form
\begin{equation}
  \alpha - \frac{\tr\alpha}{n - 1}\,\omega.
\end{equation}
\comments{  %%%%%%%%%%%%%%%
Hence, when the 2-form $\alpha = \frac{H}{n-1}\,\omega$ with $H$ a function on
$\mathcal{M}$, $\tr\alpha = \frac{n}{n-1}\,H$ and the traceless part of
$\alpha$ vanishes.
}  %%%%%%%%%%%%%%%%%%%%%%%%

If $f(q,p)$ is a function on $\mathcal{M}$, then the derivative
  $\dot{f} = \frac{\dd}{\dd t}f(q(t),p(t))$
along any one of the integral curves of
eqs.~(\ref{evps}) satisfies the equation
\begin{equation}
  \dot{f}\,\omega^n = n(n-1)\,\dd\alpha\wedge\dd f\wedge\omega^{n-2}.
\label{dotf}
\end{equation}
In fact, $\dot{f}(t) = (\Lied{X}f)(q(t),p(t))$.
And, since $\vect{X}$ is volume-preserving,
\begin{displaymath}
  (\Lied{X}f)\,\omega^n = \Lied{X}(f\,\omega^n)
  = \dd i_{\vect{X}}(f\,\omega^n) + i_{\vect{X}}\dd(f\,\omega^n)
  = \dd (f\,i_{\vect{X}}\omega^n).
\end{displaymath}
Then according to eq.~(\ref{nunX}),
\begin{eqnarray*}
  (\Lied{X}f)\,\omega^n & = & - n(n-1)\,\dd(f\,\dd\iota(\alpha))
  = - n(n-1)\,\dd f\wedge\dd(\alpha\wedge\omega^{n-2})
\\
  & = & - n(n-1)\,\dd f\wedge\dd\alpha\wedge\omega^{n-2}
  = n(n-1)\,\dd\alpha\wedge\dd f\wedge\omega^{n-2}.
\end{eqnarray*}
Thus eq.~(\ref{dotf}) has been proved.

Especially, when $\alpha = \frac{H}{n-1}\,\omega$, the system (\ref{evps})
turns out to be the usual Hamiltonian system, as we have mentioned. For such
a Hamiltonian system, on the one hand, we can use eq.~(\ref{dotf}) to obtain
\begin{displaymath}
  \dot{f}\,\omega^n = n\,\dd(H\,\omega)\wedge\dd f\wedge\omega^{n-2}
  = n\,\dd H\wedge\dd f\wedge\omega^{n-1}
  = \tr(\dd H\wedge\dd f)\,\omega^n,
\end{displaymath}
namely,
\begin{equation}
  \dot{f} = \tr(\dd H\wedge\dd f).
\end{equation}
On the other hand, $\dot{f}$ can be expressed in terms of the Poisson bracket:
\begin{displaymath}
  \dot{f} = \{f,H\} := \vect{X}_H f
  = \frac{\partial f}{\partial q^i}\frac{\partial H}{\partial p_i}
  - \frac{\partial f}{\partial p_i}\frac{\partial H}{\partial q^i}.
\end{displaymath}
So we obtain the relation between the Poisson bracket and the trace of 2-forms:
\begin{equation}
  \{f,H\} = - \tr(\dd f\wedge\dd H).
\end{equation}

In order to develop a volume-preserving
algorithm, Feng and Shang in \cite{FS} presented
a lemma as follows: The volume-preserving vector
field
   $X = (X^1, \ldots, X^n)^T$
on $\mathbb{R}^n$ can always be expressed by an
antisymmetric tensor $a^{ij}$ on $\mathbb{R}^n$
as\footnote{ In the quotation of this lemma, we
have adopted the notations so as to accommodate
ours in this letter. }
\begin{equation}
  X^i = \frac{\partial a^{ij}}{\partial x^j},
\label{divfree}
\end{equation}
where $x^i$ are the standard coordinates on
$\mathbb{R}^n$. In fact, this lemma is not
difficult to understand: A vector field on
$\mathbb{R}^n$, known as a Euclidean space or a
Riemannian manifold, is volume-preserving if and
only if it is divergence-free. Using the Hodge
theory, the divergence of the vector field $X$
reads
  $\mathrm{div} X = -\delta\tilde{X} = *\dd *\tilde{X}$,
where $\delta$ is the codifferential operator and
  $\tilde{X} = \delta_{ij}\,X^i\,\dd x^j$
is the 1-form, or the covariant vector field,
obtained from $X$ by using Einstein's method of
lowering indices. Hence that $X$ is
volume-preserving, namely, $\mathrm{div} X = *\dd
*\tilde{X} = 0$, implies $*\tilde{X}$ is a closed
$(n-1)$-form. Then, according to Poincar\'e's
lemma, there is a 2-form $\alpha$, say, such that
$*\tilde{X} = \dd *\alpha$. If $\alpha$ is
assumed to be $\frac{1}{2}\,a_{ij}\,\dd
x^i\wedge\dd x^j$ with $a_{ij} = - a_{ji} =
\delta_{ik}\delta_{jl}\,a^{kl} = a^{ij}$, we can
obtain that $\tilde{X} = (-1)^{n-1}\,*\dd *\alpha
   = \delta_{ij}\,\frac{\partial a^{jk}}{\partial x^k}\,\dd x^i$.
Then the vector field $X$ satisfies
eq.~(\ref{divfree}).

It is worthy to see that the formula
~(\ref{divfree}) is quite similar to
eq.~(\ref{XH}), only with some slight
differences: 1. This formula is a statement on
the Euclidean space $\mathbb{R}^n$ with an
arbitrary dimension $n$ while eq.~(\ref{XH}) is
for a symplectic manifold, of $2n$-dimensional.
2. It can be generalized to an $n$-dimensional
Riemannian or pseudo-Riemannian manifold provided
that the $(n-1)$-th de Rham cohomology group is
trivial, with the ordinary derivatives being
replaced by the covariant derivatives.  As for
eq.~(\ref{XH}), when
$H^{2n-1}_{\mathrm{dR}}(\mathcal{M}) \neq 0$, not
every volume-preserving vector field satisfies
it. This is similar to eq.~(\ref{divfree}). But a
covariant derivative is not necessary in
eq.~(\ref{XH}). 3. When the symplectic manifold
  $\mathcal{M} = \mathbb{R}^{2n}$,
we can set
\begin{displaymath}
  (a_{ij})_{2n\times 2n}
  = \left( \begin{array}{rcr} 0 & \ & I \\ - I & & 0 \end{array} \right)
    \left( \begin{array}{rc} Q & -A^T \\ A & P \end{array} \right)
    \left( \begin{array}{rcr} 0 & \ & - I \\ I & & 0 \end{array} \right)
  = \left(
  \begin{array}{lr}
    P   & -A \\
    A^T & Q
  \end{array}
  \right)
\end{displaymath}
where $Q = (Q_{ij})_{i=1,\, j=1}^n$, $P = (P^{ij})_{i=1,\, j=1}^n$,
$A=(A^i_j)_{i=1,\,j=1}^n$ and $I$ is the $n\times n$ unit matrix. Accordingly,
we set
  $(x^1,\ldots,x^n,x^{n+1},\ldots,x^{2n}) = (q^1,\ldots,q^n,p_1,\ldots,p_n)$.
Then eq.~(\ref{divfree}) reads
\begin{equation}
  X = \bigg(
   \frac{\partial P^{ij}}{\partial q^j} - \frac{\partial A^i_j}{\partial p_j},
   \frac{\partial A^j_i}{\partial q^j} + \frac{\partial Q_{ij}}{\partial p_j}
  \bigg)^T,
\end{equation}
with the index $i$ running form 1 to $n$. Comparing it with eq.~(\ref{XH}), it
seems that a condition
\begin{equation}
  \tr \alpha = 0
\end{equation}
could be imposed on the 2-form $\alpha$ in eq.~(\ref{XH}).

Finally, it should be mentioned that the
volume-preserving systems play an important
r\^ole in physics, especially in statistical
physics. However, this has not been noticed
widely yet. In this letter we have not explained
why the volume-preserving systems are so
important in statistical physics, but left it as
a topic for the forthcoming papers.

\vskip 1cm
\begin{center} {\Large \textbf{Acknowledgement}}
\end{center}

We would like to thank Professors J.~Z.~Pan and
Z.~J.~Shang for valuable discussions. This work
was supported in part by the National Natural
Science Foundation of China (grant Nos. 90103004,
10175070) and the National Key Project  for Basic
Research of China (G1998030601).

\end{document}